\documentstyle[12pt,epsfig]{article}
\begin{document}
\title{\bf{Equation of State of Hypermatter in $\mathbf \beta$ Equilibrium
in a Quark Model with Excluded Volume Correction}
\thanks{\it{This work was partially supported by the project
PMT-PICT0079 of ANPCYT, (FONCYT) Argentina.}}}
\author{R. M. Aguirre and A. L. De Paoli.\\
Departamento de F\'{\i}sica, Fac. de Ciencias Exactas,\\
Universidad Nacional de La Plata.\\
C. C. 67 (1900) La Plata, Argentina.}
\maketitle
\begin{abstract}

We study the effects of the finite size of the baryons on the
equation of state of homogeneous hadronic matter with hyperons.
The finite extension of hadrons is introduced in order to
improve the perfomance of field theoretical models at very high
densities, simulating the short-range hard-core repulsive part of the
baryon-baryon potential. We choose the Quark Meson Coupling model
to describe the baryon dynamics because this model provides a density
dependent radius of the bag. In addition we calculate finite size corrections
for the Zimanyi-Moszkowski model in order to make a definite comparison.
In both cases finite size effects are implemented through a
Van der Waals like correction in the normalization of the
baryon fields. In this approach we investigate $\beta$-stable  matter
in conditions that could be found in the interior of massive stars.
We find that the excluded volume corrections contribute significantly
for densities above twice the symmetric nuclear matter saturation density,
although for the quark model the results strongly depends upon the
equilibrium conditions for the confinement volume immersed in a dense
medium.\\


\end{abstract}
\newpage

\renewcommand{\theequation}{\arabic{section}.\arabic{equation}}

\section{Introduction}

Investigation of hadronic matter in extreme conditions of density and
temperature is of relevance in physics since the knowledgwe of its
properties will eventually lead to clarify some questions about
fundamental interactions and the recovering of QCD symmetries
\cite{HATSUDA,BROWN}.

The role of strangeness in the evolution of relativistic heavy-ion
collisions \cite{GREINER} as well as in the dynamics of neutron stars
\cite{BAYM} has been stated long time ago.
Near the surface of a neutron star, where the density is of the order
of the normal nuclear density $n_0$,
hadronic matter is mainly composed by protons and
neutrons in $\beta$ equilibrium with electrons, in such a way
that the whole system remains electrically neutral. At higher densities
(2-3 $n_0$) hyperons could appear, decreasing the energy per baryon.
Further increasing of the density could cause the formation of different
exotic phases composed of mesons condensates, quark matter lumps
or quark-gluon plasma.
The transition among phases are esentially pressure induced, and
differs qualitatively when there are one or two conserved charges,
{\em  e.g.} baryon number and electric charge \cite{GLEND2}.

In most hadronic matter studies, the interacting particles are
assumed as point-like.
In an early work \cite{KAPUSTA}, it was claimed the necessity to take
care of the spatial extension of the hadrons and the influence of this
fact on the collective phenomena at very high densities.
It is worthy to mention that
there are only a few field theoretical models which consistently includes
the dynamics of the hadron volume, the most commonly used are the
Skyrme and bag-like models. In the first case the inclusion of finite baryon
density effects is not straightforward due to the topological character
of its solutions \cite{RAKHIMOV}. On the other hand further refinements
of the original MIT bag model allows to deal with medium effects upon
the hadron structure \cite{Gui}-\cite{SR}.
Also in a recent work \cite{SR}, attempts to include
repulsion between overlapping bags was done through
effective short-range quark-quark correlations.

Even when finite size hadrons are considered, the associated hard core
repulsion among baryons in the nuclear medium is
frequently not taken into account. At densities lower than $n_0$
finite volume effects are small, but these repulsive
corrections could be very important at higher densities.
For the case of hyperon matter, this fact affects the
relative population of the different baryonic components. \\
The main difficulty to deal with a hard core potential is that it does
not admit a perturbative treatment at high densities. Hence it would
be desirable to have an effective treatment of these volume effects
without leaving the simplicity of the quasiparticle approximation.

We adopt a Van der Waals-like method to take into account finite volume
corrections. In this scheme the total volume appearing in the
thermodynamical quantities is replaced by the available volume,
therefore it has the advantage that the volume of the particle is
the only new parameter introduced to represent the hard core repulsive
correlations. A similar approach has been used in related investigations,
as for instance in references \cite{Waa1}-\cite{SINGH}.

In the present work we want to generalize the excluded volume
corrections to study the properties of hadron matter with strangeness,
and to check out to what extent the usual nuclear field effective
models can be used to extrapolate the hypermatter equation
of state at high densities.
We introduce these corrections at the level of the normalization
of the baryon fields. This point of view allows us to use the same
effective Lagrangians usually applied to describe the dynamics
of point-like particles. Therefore in the quasi-particle scheme of the
Mean Field Approximation for homogeneous matter, the baryons
as a whole will be described by a superposition of relativistic plane
waves, but they will move only within the accessible volume.

We shall apply this approach to study the properties of
homogeneous  hypermatter in $\beta$-equilibrium at zero temperature.
Although the formalism is stated for the whole baryon octet, in order to
clarify the discussions we shall only consider $\Lambda$ hyperons
in our calculations, besides neutrons and protons.
The equation of state of nuclear matter with lambdas has been
previously investigated in non-relativistic \cite{SCHULZE} as well as
for relativistic \cite{ZHANG} regimes considering point-like particles.
However a full treatment must include
heavier hyperons, as deduced for example, from the influence of the
$\Xi$ particles on the stability of lambdas in medium \cite{SCHAFFNER}.

In the next section we give a resume of the effective models to be used in
our calculations. Section 3 is devoted to extend these models to include
correlations based in excluded volume restrictions.
To compare the effects of the
correction proposed, we have selected two descriptions of very different
kind. On one hand we select a hadronic Lagrangian for point-like particles,
the so-called Zimanyi-Moszkowski (ZM) model \cite{ZM}, for which a
fictitious size is assigned to the baryons, and on the other
hand we have a bag model generalized to include quarks interacting with
scalar and vector mesons, the Quark Meson Coupling (QMC) model
\cite{Gui,ST}. In the last case two possible conditions for the equilibrium
of the confinement volume in dense matter are proposed. All these items
are discussed in section 4.
Numerical results and discussion are given in section 5, conclusions are
drawn in section 6.

\section{Relativistic Effective Lagrangian for Point-like Baryons}
\setcounter{equation}{0}

Relativistic Hadron Field Theories have proved to be
adequate in describing many nuclear matter properties at saturation
density, and also finite nuclei structure, with a relatively small
number of free parameters. These theories allow to extrapolate
matter behaviour to high densities and energies. Although a chiral
treatment is desirable at high densities, we do not consider this issue
in this work.

There are a variety
of effective relativistic Lagrangians, the more usual ones describe
nuclear interaction mediated by scalar ($\sigma$), vector ($\omega$)
and pseudovector ($\rho$) mesons \cite{ZM,Wa,SW}.
Since neutral matter at high densities can be composed not only by
nucleons and leptons but also with more massive baryons,
we select a Lagrangian of the form:
%(2.1)
\begin{eqnarray}
{\cal{L}}&=&\sum_B\;\left[ {\bar{\Psi}}^B ( i \gamma^{\mu}
\partial_{\mu} - g_{\omega}^B \gamma^{\mu} \omega_{\mu} - \frac
{1}{2} g_{\rho}^B \; \gamma^{\mu} {\mathbf \tau}.{\bf b}_{\mu} -
{M_B}^\ast ) \Psi^B \right] \nonumber \\ &+& \frac
{1}{2} (\partial^{\mu}\sigma \partial_{\mu}\sigma - {m_\sigma}^2
\;{\sigma}^2 ) - \frac {1}{4} F^{\mu\nu} F_{\mu\nu} + \frac {1}{2}
{m_\omega}^2\; \omega^{\mu} \omega_{\mu} \nonumber \\ &-& \frac
{1}{4} {\bf G}^{\mu\nu} {\bf G}_{\mu\nu} + \frac {1}{2}
{m_\rho}^2\; {\bf b}^{\mu}.{\bf b}_{\mu} + \sum_l\;\left[
{\bar{\Psi}}^l ( i \gamma^{\mu} \partial_{\mu} - m_l )	\Psi^l
\right]
\label{LAGRAL}
\end{eqnarray}
In Eq.(\ref{LAGRAL})  $F^{\mu\nu}$ and ${\bf G}^{\mu\nu}$ are the tensor
strength
fields of the $\omega$ and $\rho$ mesons, respectively, and the index
$B (l)$ runs over all baryons (leptons) considered. The term ${M_B}^\ast$
contains the interaction of the $\sigma$ meson with the baryonic fields
$\Psi^B$, and plays the role of the baryon effective mass.
The different forms of ${M_B}^\ast$ lead to a family of hadronic
Lagrangians. In order to have a general discussion ${M_B}^\ast$ will be
kept undetermined until needed.\\
We solve the field equations for the case of homogeneous infinite static
matter in the mean field approximation (MFA) \cite{Wa,SW}.
In this case all meson
fields are replaced by their average values, i. e.
%2.2  2.3  2.4
\begin{eqnarray}
\sigma&=&\;\;\;\sigma_0\;\;\;\;\,=-\frac{1}{{m_\sigma}^2} \sum_B
\frac{d{M_B}^\ast}{d\sigma} n_s^B
\label{SIGMA} \\
\omega_{\mu}&=&\;\omega_0 \delta_{\mu0}\;\;\,=\;\;\frac{1}{{m_\omega}^2}
\sum_B g_{\omega}^B n^B \; \delta_{\mu0} \\
b_{\mu}^a&=&b_0 \delta_{\mu0} \delta_{a3}=\;\;\frac{1}{{m_\rho}^2}
\sum_B g_{\rho}^B I_{3}^B n^B \; \delta_{\mu0} \delta_{a3}
\label{RHO}
\end{eqnarray}
where $a=1, 2, 3$ runs over all isospin directions and $I_{3}^B$
is the third isospin component of baryon $B$. In all our calculations we
use the values of $m_\sigma=550 MeV$, $m_\omega=783 MeV$, and
$m_\rho=770 MeV$ for the meson masses.\\
The equation of motion for the baryon field $B$ (spin $1/2$) is
%2.5
\begin{equation}
( i \gamma^{\mu} \partial_{\mu} - g_{\omega}^B \gamma^0 \omega_0
- g_{\rho}^B I_{3}^B\; \gamma^0 b_0 - {M_B}^\ast) \Psi^B = 0
\label{EULERLAGR}
\end{equation}
The equation above can be used to define the energy $k_0$ for a baryon
carrying  momentum $\vec{k}$
%2.6
\begin{equation}
k_0^B=\sqrt{{{M_B}^\ast}^2+{\vec{k}}^2} \pm g_{\omega}^B \omega_0 \pm
g_{\rho}^B I_{3}^B b_0
\label{PARTENER}
\end{equation}
for particle $(+)$ and antiparticle $(-)$ solutions. Within the MFA at zero
temperature only the particle solutions contributes.\\
The scalar ($n_s^B$) and baryonic  ($n^B$) densities are
defined with respect to the ground state of the system $|GS>$ for each
baryonic class
%2.7  2.8
\begin{eqnarray}
\!n_s^B \!\!\!&=& \!\!\!\!\!<GS| {\bar{\Psi}}^B \;\Psi^B
|GS>=\vartheta\;\frac{1}
{{\pi}^2} {M_ B}^\ast \int_0^{k_B} dk \frac{k^2}
{ \sqrt{{{M_B}^\ast}^2+k^2}}
\label{SCALDENS} \\
\!n^B \!\!\!&=& \!\!\!\!\!<GS| {{\Psi}^{\dag}}^B {\Psi}^B |GS> =
\;\vartheta\;
\frac {{k_B}^3} {3{\pi}^2}
\label{VECDENS}
\end{eqnarray}
In Eqs. (\ref{SCALDENS}) and (\ref{VECDENS} ) $k_B$ is the Fermi momentum
for baryons
and the factor $\vartheta = 1$ for point-like baryons.\\

For the Lagrangian (\ref{LAGRAL}) leptons behaves like free Dirac
particles.
The leptonic density $n_l$ defines the	Fermi momentum $k_l$ through
%2.9
\begin{equation}
n^l = <GS| {{\Psi}^{\dag}}^l \Psi^l |GS> = \frac {{k_l}^3} {3{\pi}^2}
\end{equation}
Mean field equations are solved for fixed total baryon density $n$ and
zero total electric density charge $n^e$:
%2.10, 2.11
\begin{eqnarray}
n&=&\sum_B n^B \\
n^e&=&\sum_B q^B n^B+\sum_l q^l n^l = 0
\label{CHARGE}
\end{eqnarray}
where $q^B$ are the baryon electric charges, $q^l=-e$ the electron
and muon charges.\\ All baryon and lepton species are assumed to
be in $\beta$ equilibrium. This constraint and the condition
(\ref{CHARGE})
%(2.11)
determine the relative population of the particles.
In the present work we consider only three baryon species,
namely $B=$ proton $(p)$,
neutron $(n)$, lambda $(\Lambda)$, and two lepton species,
$l=$ electron $(e)$, muon $(\mu)$. \\
Consequently chemical equilibrium is impossed through the following
relationships among the chemical potentials $\mu$:
%2.12
\begin{eqnarray}
\mu^n&=&\mu^p+\mu^e \nonumber \\
\mu^n&=&\mu^{\Lambda}, \;\;\;\; {\mathrm (if \; lambdas \; are \; present)}
\nonumber
\\
\mu^e&=&\mu^{\mu}, \;\;\;\; {\mathrm (if \; muons \; are \; present)}
\label{CHEMEQ}
\end{eqnarray}
with $\mu^B = k_0^B(k_B)$, see Eq. (\ref{PARTENER}),
%(2.6),
and    $\mu^l = k_0^l(k_l) =\sqrt{{m_l}^2+{k_l}^2}$.\\
One solves the system of equations (\ref{SIGMA}) to (\ref{CHEMEQ})
%(2.2) to (2.12)
for a fixed value of the total
baryon density. This determines the meson fields ($\sigma_0 ,
\omega_0, b_0$), the baryon and lepton densities ($n^B, n^l$), all the
Fermi momenta ($k_B, k_l$) and chemical potentials ($\mu^B, \mu^l$).\\
Once the solution has been found, we can calculate the total energy
density
%2.13
\begin{eqnarray}
\epsilon&=&\frac {1}{2} {m_\sigma}^2\; {\sigma_0}^2 +\frac {1}{2}
{m_\omega}^2\; {\omega_0}^2 +\frac {1}{2} {m_\rho}^2\; {b_0}^2 \nonumber \\
&+& \frac{\vartheta}{{\pi}^2} \sum_B \int_0^{k_B} dk k^2
\sqrt{{{M_B}^\ast}^2+k^2} \nonumber \\
&+&\frac{1}{{\pi}^2} \sum_l \int_0^{k_l} dk k^2 \sqrt{{m_l}^2+k^2}
\label{ENERGY}
\end{eqnarray}
and the total pressure
%2.14
\begin{eqnarray}
p&=&-\frac {1}{2} {m_\sigma}^2\; {\sigma_0}^2 +\frac {1}{2}
{m_\omega}^2\; {\omega_0}^2 +\frac {1}{2} {m_\rho}^2\; {b_0}^2 \nonumber \\
&+&\frac{\vartheta}{3{\pi}^2} \sum_B \int_0^{k_B}
\frac{dk k^4} {\sqrt{{{M_B}^\ast}^2+k^2}} \nonumber \\
&+&\frac{1}{3{\pi}^2} \sum_l \int_0^{k_l} \frac{dk k^4}
{\sqrt{{{m_l}^\ast}^2+k^2}}
\label{PRESSURE}
\end{eqnarray}

\section{A Relativistic Effective Approach for Finite Size Baryons}
\setcounter{equation}{0}

Effective Lagrangians like the one in Eq.(\ref{LAGRAL}),
describe the long range behaviour of nuclear forces. At short
distances, i.e. high densities, the strong repulsive component of
the baryon-baryon interaction appears as a consequence of the
internal structure of the particles. In turn, this short range
repulsion can be assimilated to a simplified model where baryons
are described as extended objects over a
spherical volume of radius $R$, where $R$ reflects
the range of the core repulsion.
Therefore the fraction of  available space is reduced with respect
to the case of point-like particles.
A similar approach has been applied to study the phase transition
of nuclear matter to the quark-gluon plasma \cite{Waa1,Waa2} and in
heavy-ion collisions \cite{StoGre,Cley}.\\
In our method we extend the Van der Waals prescription to take into account
the finite size of different kinds of baryons including hyperons, at any
finite density. For this purpose
we shall use the effective Lagrangian (\ref{LAGRAL}) valid
for low densities, but the solutions arising from Eq. (\ref{EULERLAGR})
%(2.5)
will be normalized within the available space.
Our approach is expected to be valid for densities such that the
center of mass of baryons be at a distance greater than $2R$ apart.
Otherwise the simple quasiparticle picture breaks down. \\
Thus we start from the plane wave solutions $\psi$ of the Dirac
equation (2.5) for fermions of spin projection $s=\pm 1/2$ carrying
momentum $\vec{k}$, and normalized in a volume $V$
%3.1
\begin{equation}
\psi^B_{\vec{k},s}(x)=V^{-1/2} u^B(\vec{k},s) e^{-ik^{\mu} x_{\mu} }
\label{FOURIER}
\end{equation}
where $u^B(\vec{k},s)$ is the spinor free particle solution in $k$
space \cite{BJORKEN}. An analogous expression holds for the
antiparticle solution.\\
We further assume that if particles have a finite size, then Eq.(3.1)
refers to the dynamics of its center of mass. Since extended hadrons are
assumed
not to overlap their motion is restricted to the available space
$V'$ defined as \cite{Waa1,Waa2}
%3.2
\begin{equation}
V'=V-\sum_B N^B v_B
\end{equation}
with $N^B$ the total number of baryons of class $B$ inside the
volume $V$, and $v_B$ is the effective volume per baryon of class $B$.
Hence if we renormalize the particle (antiparticle) wave function replacing
$V'$ for $V$,  the effective baryon fields $\Psi$ read
%3.3
\begin{eqnarray}
{\Psi}^B(x)={(V')}^{-1/2} \sum_{\vec{k},s} &[& a^{B}(\vec{k},s)
u^B(\vec{k},s) e^{-ik^{\mu} x_{\mu} } \nonumber \\
&+&b^{B \dag}(\vec{k},s) v^B(\vec{k},s) e^{\;ik^{\mu} x_{\mu} } \;]
\end{eqnarray}
as a function of the Fock space operators $a$ and $b$, for
particle and antiparticle respectively.
In this way the finite size of the baryons is
automatically accounted for into the field dynamics.\\
It is interesting to note at this point that a more detailed analysis shows that
when we have a mixture of different baryons, the excluded volume
is not exactly the same for all species, but depend upon their
respective sizes \cite{Waa3}. To simplify the discussions, in this paper we
assume that the sizes of the baryons are similar for all the species
considered (see table 1), and hence the excluded volume is, to a good
approximation, the same for every particle.\\
With respect to the effective volume per baryon $v_B$, this quantity is
expected to be proportional to the actual baryon volume, i.e. for spherical
volumes of radius $R_B$
%3.4
\begin{equation}
v_B=\alpha \frac{4 \pi}{3} {R_B}^3
\end{equation}
and for sharp rigid spheres $\alpha$ is a real number ranging from $4$,
in the low density limit, to $3\sqrt{2}/{\pi}$, which corresponds to the
maximum density allowed for non overlapping spheres, in a face
centered cubic arrange. Since we wish to apply our method to study
the high density behaviour of homogeneous isotropic matter, we
shall adopt $\alpha= 3\sqrt{2}/{\pi}$ in all our calculations.
Thus $v_B=4\sqrt{2}{R_B}^3$ and the upper limit for the baryon density
is given by $1/v_{max}=\sqrt{2}/(8{R_{max}}^3)$, where $R_{max}$
denotes the biggest radius among all present baryonic classes.\\
In order to see how volume corrections appear in our approach,
we shall use the renormalized
field (3.3) to calculate the relationship between baryon
densities $n^B$ and the Fermi momenta $k_B$\\
%3.5
\begin{equation}
\label{DENSREN}
n^B = V^{-1} \int_V dx^3 <GS| {{\Psi}^{\dag}}^B {\Psi}^B |GS>=
(1-\sum_{B'} n^{B'} v_{B'})\;\frac {{k_B}^3} {3{\pi}^2}
\end{equation}
where $\sum\limits_{\vec{k}} \rightarrow V'/(2\pi^3) \int dk^3$ has been
used.\\
This result is equivalent to Eq. (\ref{VECDENS}),
%(2.8),
if the factor $\vartheta$ takes now the value
%3.6
\begin{equation}
\vartheta=1-\sum_{B} n^{B} v_{B}
\label{THETA}
\end{equation}
for the case of finite baryon effective volumes $v_B$ . In the
limit $v_B \rightarrow 0 $ one recovers the point-like
expressions.\\
Thus we can still use the same expressions we have
calculated in section 2.1 with $\vartheta$ satisfying Eq. (\ref{THETA}) \\
%(3.6).\\
Equation (\ref{DENSREN})
%(3.5)
shows that volume correlations couples non linearly the baryons among
themselves, in a density dependent way. \\
We can solve explicitly Eq. (\ref{DENSREN})
%(3.5)
for $n_B$ as a function of all the Fermi momenta $k_B$, namely
%3.7
\begin{equation}
n^B=\frac{1}{(1+\sum\limits_{B'}\frac{{k_{B'}}^3}{3{\pi}^2}\; v_{B'})}
\;\frac{{k_B}^3}{3{\pi}^2}
\end{equation}
Since $\vartheta$ in our approach depends explicitly upon the
baryonic densities,  the chemical potentials get an
extra term, i.e.
%3.8
\begin{eqnarray}
\mu^B&=&{ \left( \frac{\partial\epsilon}{\partial n_B} \right) }_
{{ n_{{B'}_{{}_{{}_{{}_{\!\!\!\!\!\!\!\!\!\!\!B' \neq B}}}}}}} =
\mu_0^B + \Delta \mu^B
\label{PQ}
\end{eqnarray}
where $\mu_0^B=k_0^B (k_B)$, Eq. (\ref{PARTENER}),
%(2.6),
is the free quasiparticle eigenvalue and
%3.9
\begin{equation}
\Delta \mu^B =\frac{v_B }{3{\pi}^2} \sum_{B'} \int_0^{k_{B'}}
\frac{dk k^4} {\sqrt{{{M_{B'}}^\ast}^2+k^2}}
\label{POTQUIM2}
\end{equation}
This value for $\Delta \mu^B$ is in agreement with the one used in the
literature \cite{Waa1} - \cite{Cley}.\\
The expression for the energy density $\epsilon$ is formally the same
given in (\ref{ENERGY}),
%(2.13),
and the total pressure $p$ adquires an additional term $\Delta p$
%3.9
\begin{eqnarray}
p=p_0 + \Delta p = p_0 + \sum\limits_{B}  n^B \Delta \mu^B
\label{NEWPRESS}
\end{eqnarray}
where $p_0$ is defined in Eq. (\ref{PRESSURE}),
%(2.13),
and $\vartheta$ is evaluated according to Eq. (\ref{THETA})
%(3.6)
in all the expressions.\\

In our approach only the baryonic states
receive an explicit correction due to short range forces, through the
normalization of the fields $\Psi^B$.
This can be justified because meson fields
contribute only through their mean values, which are completely
determined from the baryonic sources. These sources
already include finite size corrections in $\vartheta$.
Leptons do not experiment strong interactions, and in
this sense they are taken as point-like particles.
To resume, in our approach to finite volume correlations, matter
properties are determined applying the set of equations (\ref{SIGMA}) to
%(2.2) to (2.14)
(\ref{ENERGY}), together with (\ref{PQ})-(\ref{NEWPRESS}) for a
 fixed value of the total baryon density
$n$, but using the value of $\vartheta$ given in (\ref{THETA}) instead.\\

\renewcommand{\theequation}
{\arabic{section}.\arabic{subsection}.\arabic{equation}}

\section{Effective Baryonic Models}
\subsection{The Zimanyi-Moszkowski Model}
\setcounter{equation}{0}

In order to have definite predictions of finite size effects we
apply the scheme described in section 3, to the ZM model \cite{ZM}.
This is a local field model which basic degrees of freedom are
point-like hadrons. This
model has been extensively used to study nuclear matter and nuclei
properties.
It has the advantage of reproduce the main nuclear matter properties
in the lowest order of approximation with reasonable accuracy, and
using the minimum set of free parameters.
In the ZM model, properly generalized to include several baryonic species,
the
effective baryon mass ${M_B}^\ast$ and its
derivative $d{M_B}^\ast/d\sigma$ are given by
%4.1.1
\begin{eqnarray}
{M_B}^\ast&=& M_B {(1+g_{\sigma}^B \sigma /M_B)}^{-1} , \nonumber \\
\frac{d{M_B}^\ast}{d\sigma}&=&-g_{\sigma}^B
{(1+g_{\sigma}^B \sigma /M_B)}^{-2}
\end{eqnarray}
respectively.\\
We  place this expression for  ${M_B}^\ast$ in the equations of
sections 2 and 3,
to calculate the effects of finite size corrections in the thermodinamical
bulk properties.\\
If we take $\vartheta=1$, section 2, we describe a system of point-like
baryons, and using $\vartheta$ from Eq. (\ref{THETA}),
%(3.6),
we include finite size effects.
In the last case a fictitious volume of constant radius is assigned to
each kind of baryons in order to contrast excluded volume effects with
the point-like case. However, the high density results must be taken
with caution since causality violation is expected due to the hard-core
repulsion \cite{StoGre}. This problem can be avoided if the
baryon volume decreases with density as predicted by the model presented in
the next subsection.\\
Numerical results will be given in section 5.

\subsection{The Quark Meson Coupling Model}
\setcounter{equation}{0}

Another baryon model we shall use is the Quark-Meson Coupling (QMC) model
\cite{Gui,ST}. The QMC model may be
viewed as an extension of the Quantum Hadro-Dynamics (QHD) models
like, for instance, the Walecka \cite{Wa,SW} or the
Zimanyi-Moszkowski \cite{ZM} models. In the QMC model hadrons are
represented as non-overlapping bags containing three valence quarks, the
bag radius changing dinamically with the medium density. Baryons are
interacting by the exchange of $\sigma$ and
$\omega$ mesons that couples directly to the
confined quarks. It has been found that
these extra degrees of freedom provided by the internal structure
of the baryon lead to quite acceptable values of the nuclear
matter compressibility at saturation. Despite the explicit
quark fields present in the QMC model, hadronic thermodynamical
properties are evaluated in such a way that baryons are handled as
point-like particles with an effective mass $M_B^{\ast}$ which
depends on the $\sigma$ field.\\
To describe briefly the QMC model, one considers baryons as
spherical MIT bags where quarks are confined, the
Dirac equation for a quark of flavor $q, \; (q=u, d, s)$, of current mass
$m_q$ and $I_{3}^q$ third isospin component is then given by
%4.2.1
\begin{equation}
( i \gamma^{\mu} \partial_{\mu} - g_{\omega}^q \gamma^0 \omega_0 -
g_{\rho}^q I_{3}^q\; \gamma^0 b_0 - {m_q}^\ast) \Psi^q = 0
\label{QMCEQ}
\end{equation}
In this equation all meson fields have been replaced by their
respective mean field values, in a similar way as we have done in
Eq.(\ref{EULERLAGR}) for baryons in uniform hadronic matter.
Mesons are supposed to couple linearly only to
$u$ and $d$ quarks, i.e. $g_{\sigma}^s=g_{\omega}^s=g_{\rho}^s=0$.
Therefore the effective ${m_q}^{\ast}$ masses are
%4.2.2
\begin{eqnarray}
{m_{u,d}}^\ast&=&m_{u,d} - g_{\sigma}^{u,d} \sigma \nonumber \\
{m_s}^{\ast}&=&m_s
\label{QMASS}
\end{eqnarray}
For a spherically symmetric bag of radius $R_B$, representing a baryon of
class $B$, the normalized quark wave
functions $\Psi^q_B(r,t)$ are given by
%4.2.3
\begin{equation}
\Psi^q_B(r,t)={\cal N}_B^{-1/2} \frac{e^{-i{\varepsilon}_{q B} t}}
{\sqrt{4\pi}} \left( \begin{array}{c}
j_0 (x_{qB} \, r/R_B) \\
i \beta_{q B} {\vec{\sigma}}.{\hat{r}} j_1 (x_{q B} \, r/R_B)
\end{array} \right) \chi ^q
\end{equation}
where $\chi ^q $ is the quark spinor and
%4.2.4
\begin{equation}
\varepsilon_q = \frac{\Omega_{q B}}{\!R_B} + g_{\omega}^q \;\omega_0 +
g_{\rho}^q I_{3}^q\; b_0
\end{equation}
%4.2.5
\begin{equation}
{\cal N }_B={R_B}^3\;[2 \Omega_{q B} (\Omega_{q B} - 1) +
R_B {m_q}^\ast ]\; \frac{ j_0^2 (x_{q B}) }{x_{q B}^2}
\end{equation}
%4.2.6
\begin{equation}
\beta_{q B} ={\left[ \frac{\Omega_{q B} - R_B {m_q}^\ast }{\Omega_{q B} +
R_B {m_q}^\ast } \right]}^{1/2}
\end{equation}
with $\Omega_{q B} =[x_{q B}^2 +{(R_B {m_q}^\ast)}^2	]^{1/2}$.
The eigenvalue $x_{q B}$ is solution of the equation
%4.2.7
\begin{equation}
j_0 (x_{q B}) = \beta_q \; j_1 (x_{q B})
\label{BOUNDARY}
\end{equation}
which arises from the boundary condition at the surface of the bag.\\
In this model the ground state bag energy is identified with the baryon
mass ${M_{B}}^\ast$,
%4.2.8
\begin{equation}
{M_{B}}^\ast=\frac{\sum_q n_q^B \Omega_{q B} - z_{0 B}}{R_B} +
\frac{4}{3} \pi B_0 {R_B}^3
\label{BAGMASS}
\end{equation}
where $n_q^B$ is the number of quarks of flavor $q$ inside the bag,
$B_0^{1/4} = 210.85 MeV$ is the bag constant and $z_{0 B}$
the zero-point motion parameter, fixed to reproduce the baryon spectrum
at zero density.\\

\begin{table}[ht]
\centering

\begin{tabular}{|l|c|c|c|} \hline
baryon & $M (MeV/c^2)$ & $R (fm)$ & $z_0$ \\ \hline
  p		   & 938.2723 & 0.6 & 4.00496464 \\ \hline
  n		   & 939.5656 & 0.6 & 4.00102120 \\ \hline
  $\Lambda$ & 1115.63 & 0.606 & 3.86577814 \\ \hline
\end{tabular}

\caption{\footnotesize{Baryon parameters. $M$ and $R$ are, respectively,
the mass
and the radius at zero density. For the ZM model $R$ is density independent.
The bag constant $B_0$ has been fixed by the value $B^{1/4}=210.8 MeV$.}}
\end{table}
To obtain this mass one has first to solve Eq. (\ref{BOUNDARY}), with
(\ref{QMASS}) and the $\sigma_0$ mean field given by (\ref{SIGMA})
where the derivative in this equation must be interpreted as
$\left(\partial M^{\ast}_B/ \partial \sigma \right)_{R}$. \\
Equation (\ref{BAGMASS}) shows that the baryon effective mass is a
function of the bag radius $R_B$. In the original MIT bag calculations
$R_B$ is a constant fixed at zero baryon density, but in the QMC it is
a variable dynamically adjusted to reach the equilibrium of the bag
in the dense hadronic medium.
We shall consider two possibles equilibrium conditions, to be combined
with the point-like and finite size considerations. They are:\\

a) the standard QMC prescription which minimizes
${M_{B}}^\ast$ with respect to $R_B$ keeping the $\sigma$ field constant,
namely \cite{ST}
%4.2.9
\begin{equation}
{\left( \frac {\partial {M_{B}}^{\ast}} {\partial R_B} \right)}_{\sigma}=0
\end{equation}
This prescription will be denoted as $QMCa$.

b) an alternative version of the QMC model, more suitable
to account for the bag equilibrium at high pressures, is obtained by
impossing that the net momentum flux through the bag surface be zero
\cite {AS} (see Appendix)
%4.2.10
\begin{equation}
- {\left( \frac
{\partial {M_B}^\ast} {\partial v_B} \right)}_{\sigma, x_{q B}}
=\frac{1}{3{\pi}^2} \sum_{B'} \int_0^{k_{B'}}
\frac{dk k^4} {\sqrt{{{M_{B'}}^\ast}^2+k^2}}
\label{QMCb}
\end{equation}
where the left hand side is the pressure exerted by the quarks inside
the bag of kind $B$ and the right hand side is the baryon contribution
to the total hadronic pressure.\\
This condition will be labelled as $QMCb$ in what follows.

It must be mentioned that a similar condition can be obtained
by minimizing the energy density $\epsilon$ with respect to
the effective bag volume $v_B$, keeping constant the total volume
$V$, namely \cite{Waa1,Waa2}
%4.2.11
\begin{equation}
-\frac {n^B_s}{n^B} {\left( \frac
{\partial {M_B}^\ast} {\partial v_B} \right)}_{\sigma, x_{q B}} =
 \frac{1}{3{\pi}^2} \sum_{B'} \int_0^{k_{B'}}
\frac{dk k^4} {\sqrt{{{M_{B'}}^\ast}^2+k^2}}
\label{QMCc}
\end{equation}
The only difference with Eq.(\ref{QMCb}) stands in the factor
${n^B_s}/{n^B}$. This factor is found numerically to be close to unity in
the range of densities we handle in the present work. In fact, numerical results
using condition (\ref{QMCc}) differ only by negligible amounts with respect
to the predictions arising from (\ref{QMCb}) for symmetric nuclear matter.\\
Hence we prefer to use the much simpler condition (\ref{QMCb}) for further
calculations.

Conditions $QMCa$ and $QMCb$ coincide only in the case of vanishing
density, but the last one seems to be more appropriate to describe general
equilibrium situations of bags inmersed in high density medium.\\

Once ${M_{B}}^\ast$ has been defined microscopically, the dynamics of
the hadrons in the QMC model arises from the effective Lagrangian
(\ref{LAGRAL}),
as discussed in section 2.\\
Only the couplings of mesons to baryons remains to be properly defined,
since
in this model mesons interacts directly with quarks. In fact,
the quark-meson couplings $g_{\phi}^q$, with $\phi = \sigma, \omega,
\rho $ and $q= u,d$, are related to
the corresponding meson-baryon couplings in a simple way
\cite{ST}.
If we denote with $g_{\phi}^{ B}$ the coupling of the $\phi$-meson to
$B$-baryon, and
calling $n^B_{ns}$ the non-strange quark content of baryon $B$, then
%4.2.12
\begin{eqnarray}
g_{\sigma}^{ B} &=& \frac {n^B_{ns}} {3} g_{\sigma}^{u} \nonumber \\
g_{\omega}^{ B} &=& \frac {n^B_{ns}} {3} g_{\omega}^{u}  \\
g_{\rho}^{ B} &=& g_{\rho}^{u} \nonumber
\end{eqnarray}

When $\vartheta=1$ (see section 2), we neglect the particle size
in the evaluation of many-body effects. Otherwise these are included
using Eq.(\ref{THETA}) for $\vartheta$ instead.\\
An interesting fact related to the QMC model is that the bag volume
reflects the changes in the medium density
through the mean value of the $\sigma$ field appearing in either of
the equilibrium conditions $QMCa$ or $QMCb$.
This is a
self-consistency condition since $\sigma$ itself depends on the effective
volume of the baryons through the coefficient $\vartheta$. \\
Numerical results will be discussed in the next section.

\begin{table}[ht]
\centering

\begin{tabular}{|l|c|c|c|c|c|c|} \hline
Model & Case &	$g_{\sigma}$ & $g_{\omega}$ &  $g_{\rho}$ &
$g_{\sigma}^{\Lambda}$ & $g_{\omega}^{\Lambda}$ \\ \hline
 ZM	& $NC$	& 7.8449 & 6.6710 & 9.4802 & 4.6760 & 4.9620 \\ \cline{2-
7}
	   & $VC$  & 7.4039 & 5.8275 & 9.2446 & 4.1997 & 4.2872 \\ \hline
$QMCa$   & $NC$ & 17.964 & 9.012 & 9.220 & 11.977 & 6.008 \\ \cline{2-7}
 & $VC$ & 17.007 & 8.172 & 8.873 & 11.338 & 5.448 \\ \hline
$QMCb$   & $NC$ & 17.985 & 9.024 & 9.219 & 11.990 & 6.016 \\ \cline{2-7}
 & $VC$ & 17.046 & 8.199 & 8.870 & 11.364 & 5.466 \\ \hline
\end{tabular}

\caption{\footnotesize{Baryon-meson couplings $g_{\sigma, \omega, \rho}$ 
used in our
calculations, for the ZM, $QMCa$ and $QMCb$ models as described in the text,
and combined with the point-like ($NC$) and finite size ($VC$) cases. The
couplings has been adjusted in order to reproduce the symmetric nuclear matter
properties at saturation in either case,  see section 5.1.}}
\end{table}

\section{Finite size effects: Numerical Results}

\subsection{Homogeneous symmetric nuclear matter}
\setcounter{equation}{0}

Nuclear matter properties has been extensively studied using point-like
hadronic models. We wish to investigate excluded volume corrections
in the homogeneous symmetric nuclear matter equation of state (EOS).
With this purpose we shall use alternatively the hadronic models ZM and
QMC to apply the scheme described in section 3. Physical predictions for
 point-like ($\vartheta=1$) and finite extension baryons will be
compared, and for the QMC case we also have considered two
alternative conditions for the equilibrium of bags in the nuclear medium.
In the following we shall use the labels $VC$ and $NC$ to indicate results
obtained with or without excluded volume correction, respectively. Also
the labels $a$ and $b$ will be associated with the QMC model to
distinguish between the proposed equilibrium conditions, as explained in
the previous section. \\

\begin{figure}[ht]
\begin{tabular}{cc}
\begin{minipage}[t]{6.6cm}
\epsfig{file=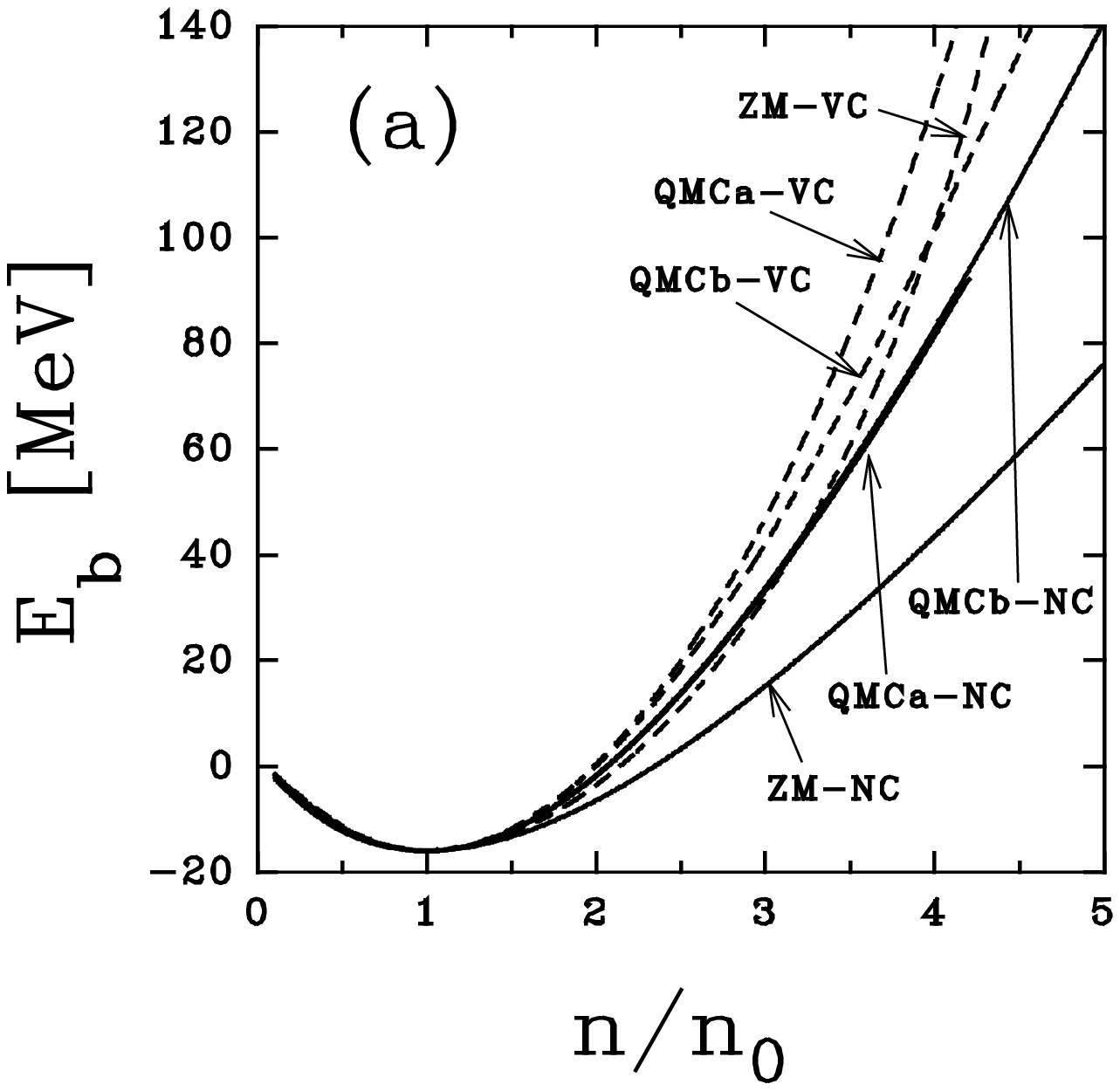,width=1.\textwidth}   
\end{minipage} &
\begin{minipage}[t]{6.6cm}
\epsfig{file=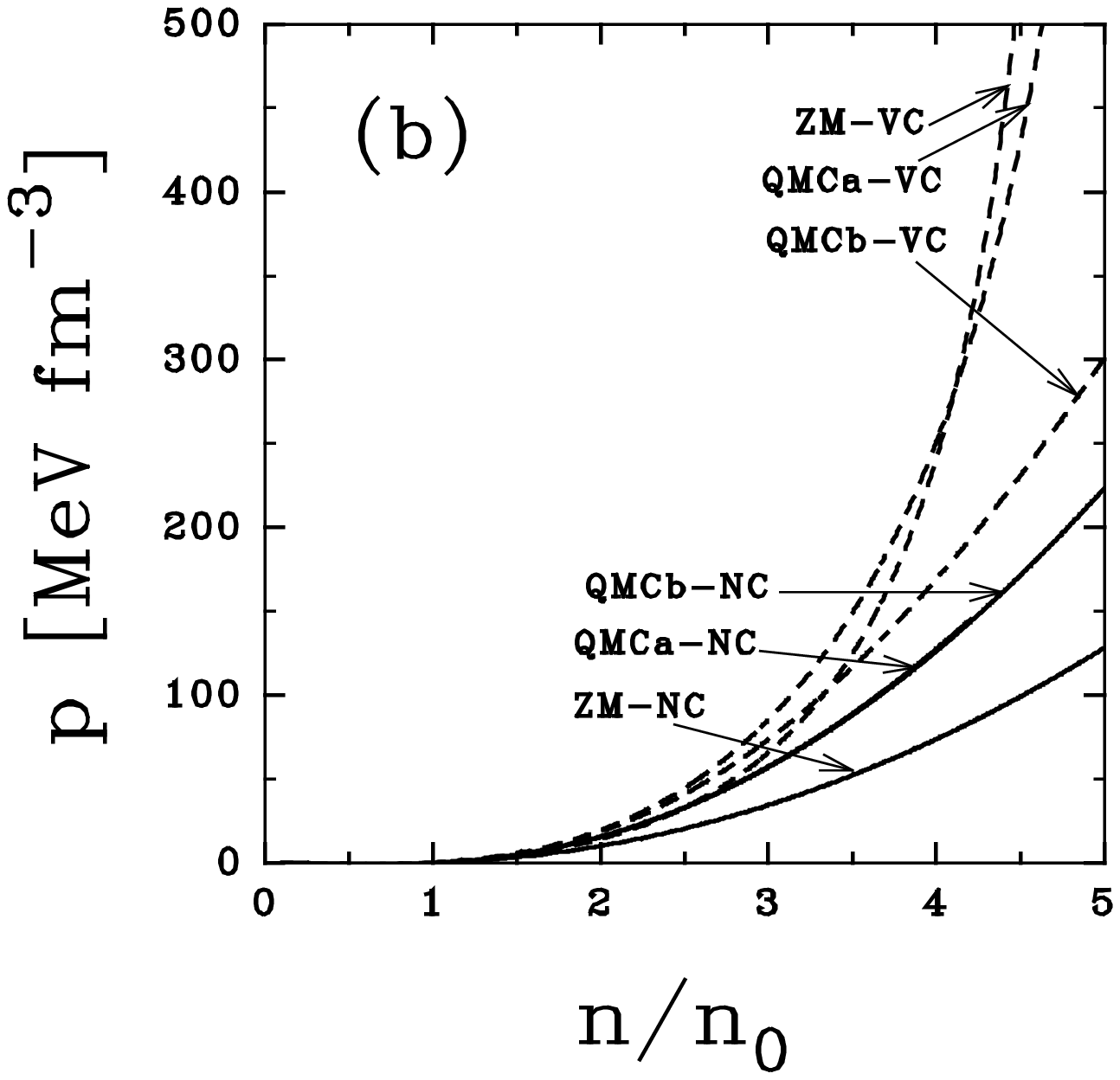,width=1.\textwidth}
\end{minipage}
\end{tabular}
\caption{\footnotesize{Thermodynamical properties of symmetric nuclear matter:
binding
energy $E_b = (\epsilon/n) -\bar{M} c^2$ (a) and pressure (b) as functions
of the relative baryon density. The curves corresponds to the models
Zimanyi-Moszkowski (ZM)
and Quark-Meson Coupling (QMC). Results with excluded volume correction
($VC$) are represented with dashed lines and those without this correction
($NC$) by solid lines. For the QMC model the cases $QMCa$ and $QMCb$
arise by selecting different equilibrium conditions as explained in the text.}}
\end{figure}

The equations to be solved for nuclear matter are the same as those in
the neutral matter case described in section 2, except that
proton and neutron densities are equal ($n^p=n^n$),
hyperons and leptons are absent and hence
Eq.(\ref{CHARGE}) for electrical neutrality no longer holds.\\
In tables 1 and 2 we display the set of parameters to be used in both
the ZM and QMC models, respectively, combined with different
possibilities, as explained in the table captions.
Each set of parameters has been obtained with the constraint that
the known
symmetric nuclear matter properties at saturation must be reproduced, i.e.
the saturation density $n_0$, the binding energy per nucleon
${E_b}_0$, and the symmetry energy coefficient $a_s$:
%5.1.1
\begin{eqnarray}
& & n_0 = \;0.15 fm^{-3}  \nonumber \\
& & {E_b}_0 = {(\epsilon/n)}_0 -\bar{M} c^2 = -16 MeV  \nonumber \\
& & a_s = \frac{1}{2} {\left( \frac{\partial^2 (\epsilon/n)}{{\partial t}^2}
\right)}_{t=0} =\;35 MeV
\end{eqnarray}
where $n=(n_n + n_p)$ is the total nucleon density, $t=(n_n - n_p)/n$ and
$\bar{M} = 938.92 MeV/{c^2}$ is the average free nucleon rest mass.\\
When finite size baryons are considered we fix the nucleon radius
in vaccum at $0.6 fm$; with this choise the upper nuclear
density allowed for non overlapping bags is always
beyond $n/{n_0} = 5$, for the set of parameters in tables 1 and 2. \\
In figure 1a we plot the binding energy $E_b =(\epsilon/n) - \bar{M} c^2$, as
function of the relative baryon density $n/{n_0}$ up to $5$ times the
saturation density, for both ZM and QMC models, with and without finite
volume corrections. Also we show in figure 1b the total pressure
for symmetric nuclear matter, corresponding to the same
cases depicted in figure 1a.
We can see in either figures 1a and 1b that all the cases practically
coincide in the range $0 < n/{n_0} < 1.5$. Thus, as expected,
in this density domain the short range
nucleon-nucleon repulsion can be neglected. \\
In particular the nuclear matter compressibility at saturation,
$\kappa = 9 {(\partial p / \partial n)}_{n_0}$, is only slightly increased
by finite volume corrections, the increment is about $15 \%$ in the ZM
and about $8 \%$ in the QMC models, see table 3.\\

\begin{table}[ht]
\centering

\begin{tabular}{|l|c|c|c|} \hline
Model & Case & ${M_{p}}^\ast/M_p$ & $\kappa$ (MeV) \\ \hline
 ZM	       & $NC$  & 0.850 & 221.4 \\ \cline{2-4}
		  & $VC$  & 0.862 & 254.1 \\ \hline
$QMCa$ & $NC$ & 0.773 & 300.6 \\ \cline{2-4}
		  & $VC$  & 0.792  & 323.4 \\ \hline
$QMCb$ & $NC$  & 0.772 & 296.1 \\ \cline{2-4}
		  & $VC$  & 0.791 & 318.6 \\ \hline
\end{tabular}

\caption{\footnotesize{Predicted properties for symmetric nuclear matter at 
saturation,
using the ZM and QMC models with the corresponding set of
couplings constants detailed in Table 2.}}
\end{table}

For the QMC formalism the cases $a$ and $b$ are practically
undistinguishable when volume corrections are not included,
meanwhile in the $VC$ approach there are significative differences only
for densities higher than $2.5 n_0$.
For every case the inclusion of the proposed corrections enhances, for
a given density, both pressure and binding energy. This increment is
greater for $QMCa$ than for $QMCb$. This behaviour can be
explained by looking at figure 2a, where we plot the
density dependence of the proton radius $R_p$.
In fact, one sees there that all predictions are very similar for densities
below $n_0$, but at higher densities the bag size decreases much faster in
the option $QMCb$ including or not volume corrections, as compared with the
option $QMCa$ and therefore in the option $QMCb$ $\vartheta$ is always
closer to unity. The behaviour shown in figure 2a could be expected because
in the condition $QMCb$ the bag must support the growing
external baryon pressure, while option $QMCa$ only equilibrates the
in-medium bag with its internal structure.\\
To conclude, in case $QMCb$ the excluded volume remains negligible in a
wider density range, allowing the baryonic system to compress
to much higher densities as compared with the standard option $QMCa$,
or with the rigid, undeformable baryon used for the ZM model.\\
Also it must be noted in this figure that corrections effects have opposite
signs in cases $QMCa$ and $QMCb$.\\
The pronounced compression of the bags predicted by the $QMCb$
case could justify the use of effective point-like equations to
extrapolate thermodynamical bulk properties at high densities, provided the
bag compression is appropriately accounted for in the effective baryon
mass.\\
\begin{figure}[ht]
\begin{tabular}{cc}
\begin{minipage}[t]{6.6cm}
\epsfig{file=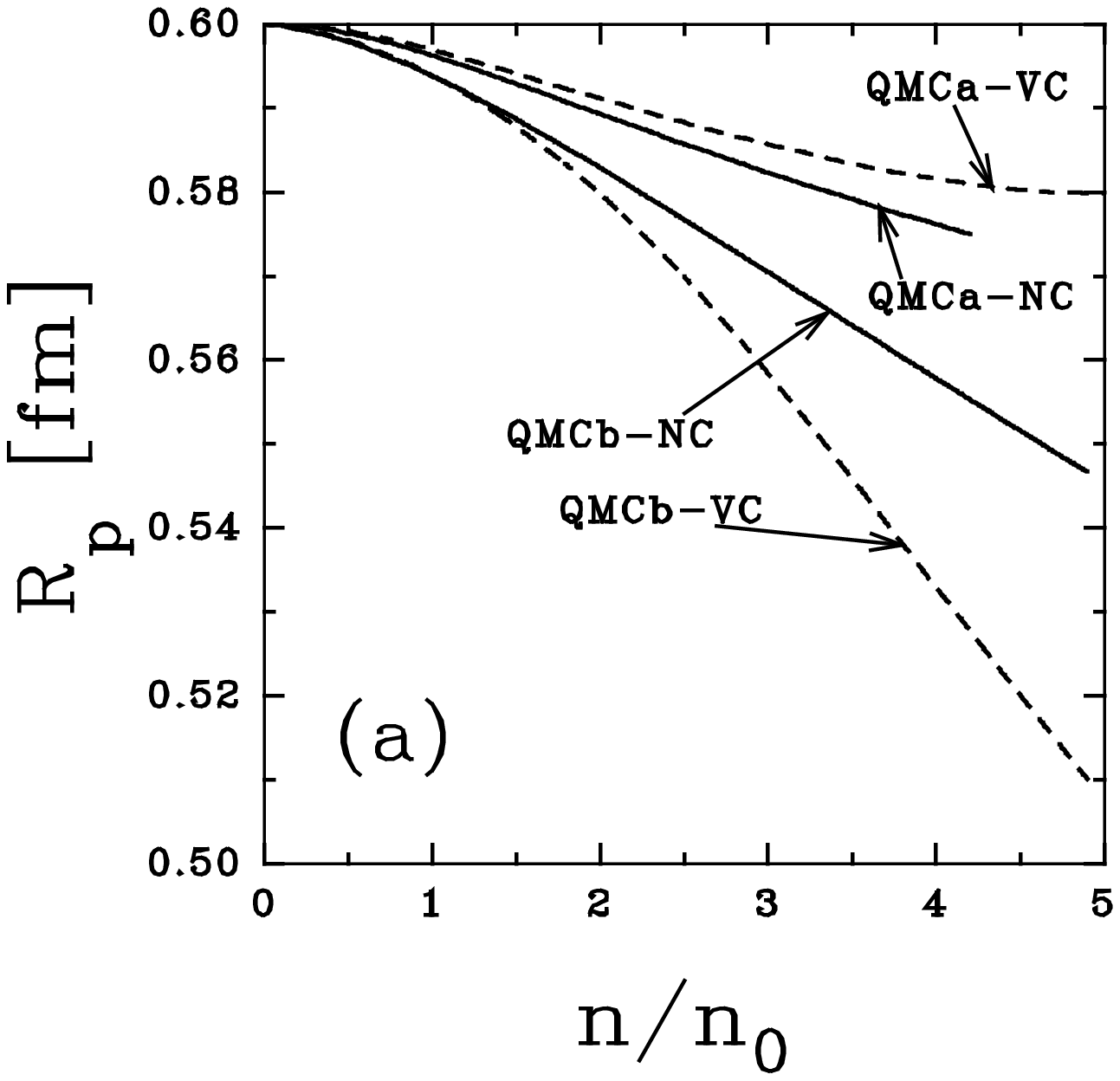,width=1.\textwidth}   
\end{minipage} &
\begin{minipage}[t]{6.6cm}
\epsfig{file=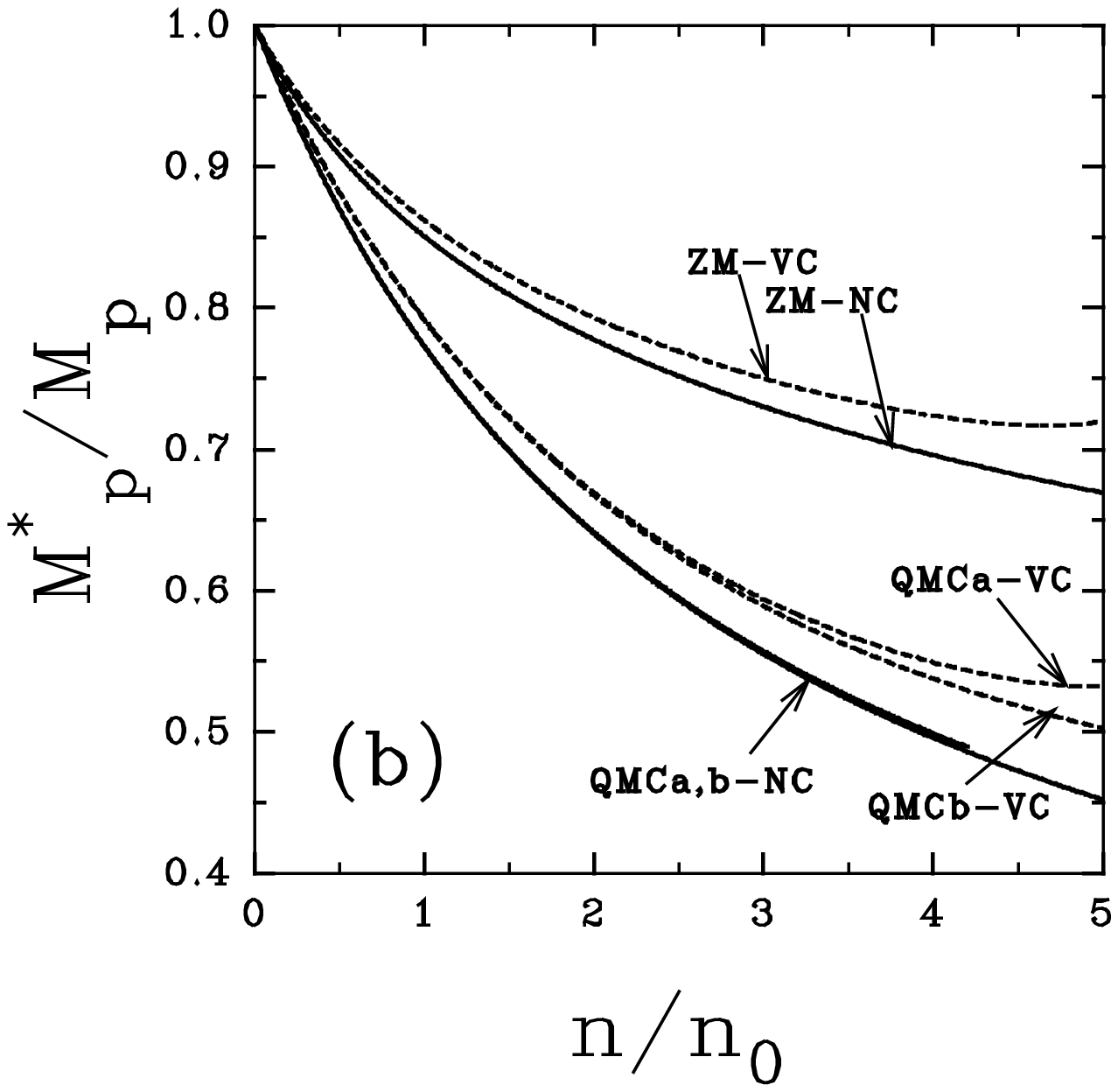,width=1.\textwidth}
\end{minipage}
\end{tabular}
\caption{\footnotesize{
Properties of nucleon immersed in symmetric nuclear matter: the
proton radius (a) and the proton mass (b) as functions of the relative baryon
density. In the first case only QMC results are shown since in the modified
ZM model a constant radius is assumed for all the baryons. The lines and
labels conventions are the same as those defined in fig. 1.}}
\end{figure}

In figure 2b we plot the behaviour of the proton effective mass
${M}^\ast_p /M_p$ as function of the relative nucleon density. This figure
shows that repulsive finite size effects tends to stabilize the effective mass
when the density increases, for both the ZM and QMC models. \\
On the other hand the difference between equilibrium conditions $QMCa$
and $QMCb$ becomes significative for the effective mass only for extremely
high densities.

\subsection{Homogeneous neutral matter}
\setcounter{equation}{0}

We proceed to calculate numerically the effects of finite baryonic size on
neutral matter properties, using alternatively the ZM and QMC  models.
The parameters inherent to each model
have been fixed in the previous section, tables 1 and 2.\\

\begin{figure}[ht]
\begin{tabular}{cc}
\begin{minipage}[t]{6.6cm}
\epsfig{file=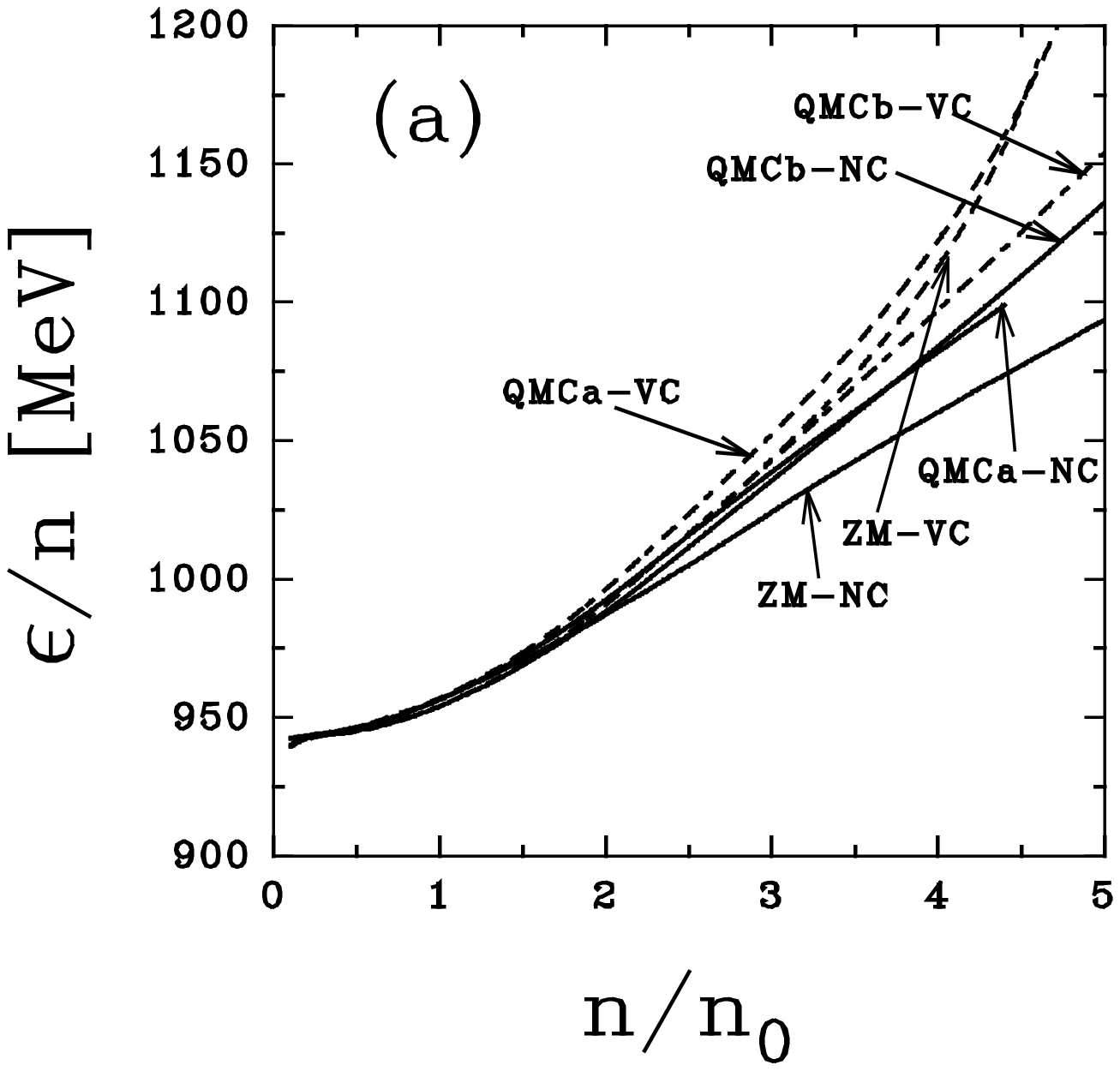,width=1.\textwidth}   
\end{minipage} &
\begin{minipage}[t]{6.6cm}
\epsfig{file=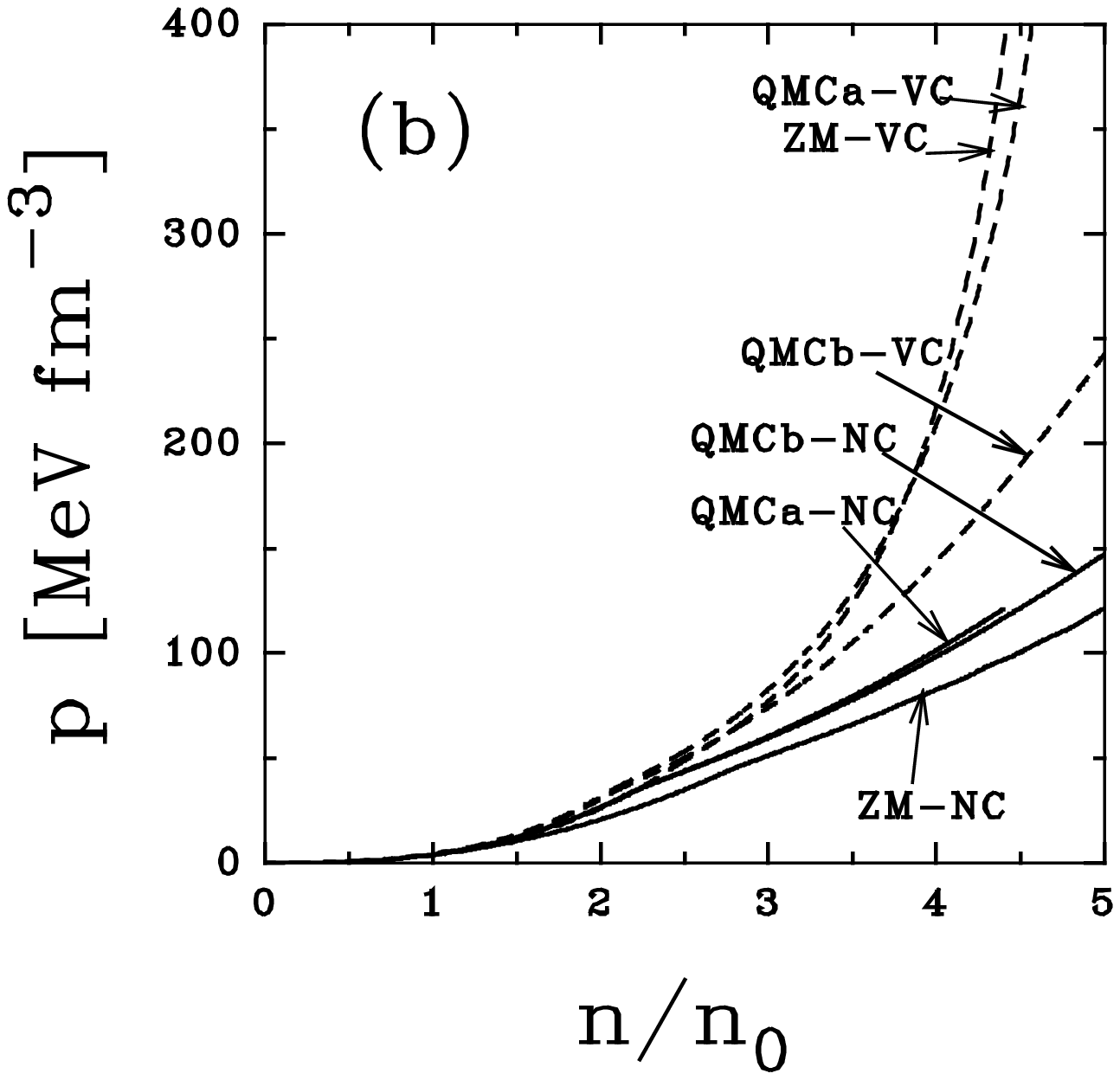,width=1.\textwidth}
\end{minipage}
\end{tabular}
\caption{\footnotesize{
Thermodynamical properties of electrically neutral nuclear matter
with $\Lambda$ hyperons in $\beta$ equilibrium with leptons: the energy per
particle (a) and the pressure (b) as functions of the relative baryon density.
The lines and labels conventions are the same as those in fig. 1.}}
\end{figure}

To simplify the discussion, we study hyperon neutral matter compossed
only by protons, neutrons and lambdas, together with electrons and muons.\\
In the ZM case the coupling of the hyperon $\Lambda$ to mesons must
be determined from experimental data. We use the condition that
the binding energy of a single $\Lambda$ in symmetric nuclear matter must
have a minimum of $-28 MeV$ \cite{MD} at the saturation density $n_0$.
This condition determines the values $g_{\sigma, \omega}^{\Lambda}$,
which are quoted in table 2, for the $NC$ and $VC$ cases.
In the QMC framework the $\Lambda$ coupling constants are reduced to
$2/3$ of the value corresponding to non-strange nucleons; numerical values
of $g_{\sigma, \omega}^{\Lambda}$ are quoted in table 2.

In figures 3a and 3b the relativistic energy per baryon $\epsilon/n$ and the
pressure $p$ are drawn versus the relative baryon density, respectively.
Most of the general conclusions given for nuclear matter can be repeated
here:
the $VC$ approach increases, for a given density both pressure and energy,
independently of the model used.\\

\begin{figure}[ht]
\begin{tabular}{cc}
\begin{minipage}[t]{6.6cm}
\epsfig{file=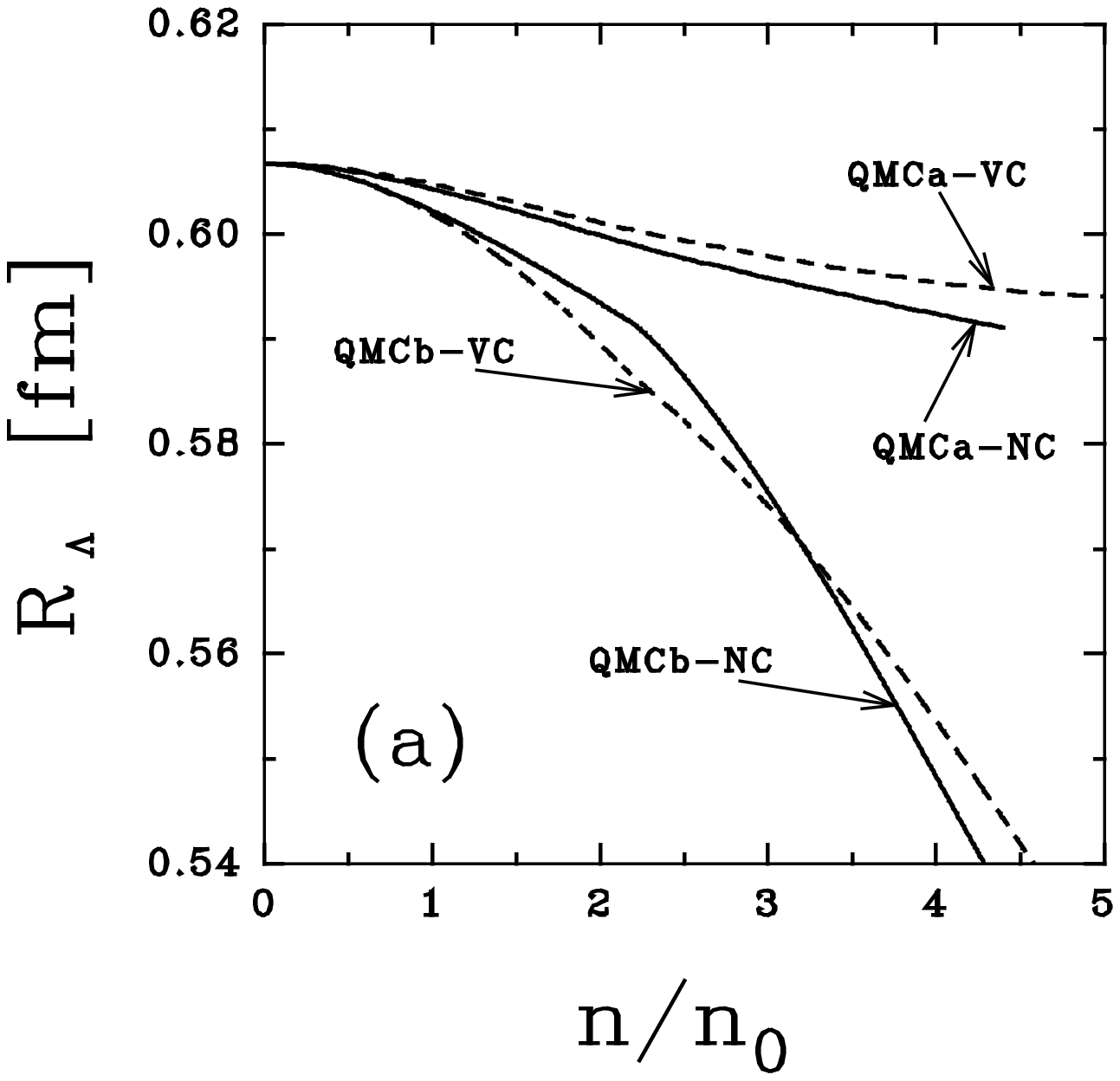,width=1.\textwidth}   
\end{minipage} &
\begin{minipage}[t]{6.6cm}
\epsfig{file=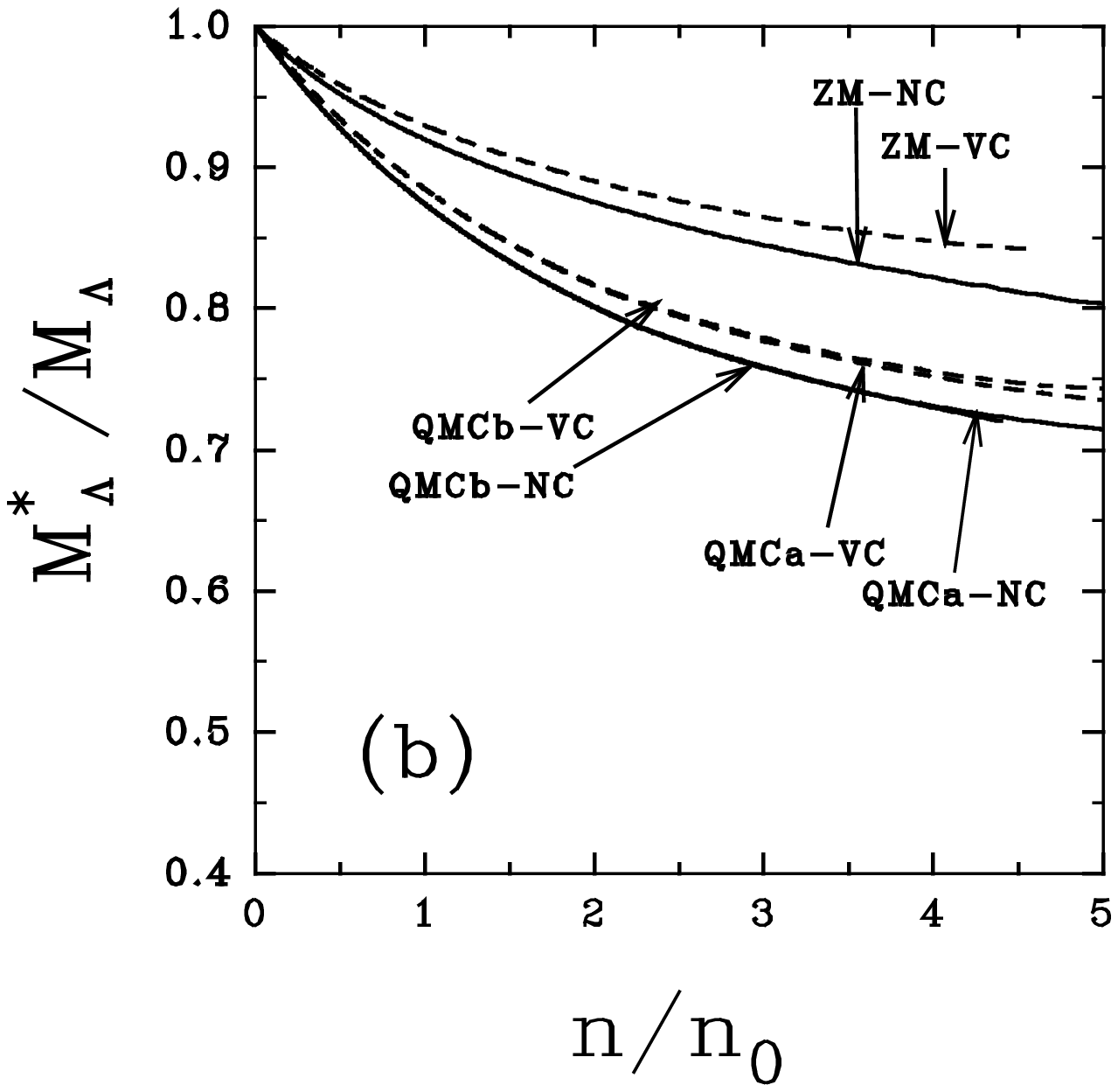,width=1.\textwidth}
\end{minipage}
\end{tabular}
\caption{\footnotesize{
Properties of the $\Lambda$ in neutral nuclear matter in $\beta$
equilibrium: the bag radius (a) and the effective mass (b) are shown as
functions of the relative baryon density. In the first case only QMC results
are displayed. The lines and labels conventions are the same as those in fig. 1.
}}
\end{figure}

The effects of this correction are more significative for ZM and $QMCa$
and less important for $QMCb$.\\
The density dependence of the raddii and effective masses of nucleons are
very similar to the result obtained for symmetric nuclear matter.
The behaviour of $R_{\Lambda}$ with the density is depicted in
figure 4a for the QMC model, and the density dependence of
the $\Lambda$ mass appears in figure 4b; it can be seen that finite volume
corrections enhance this effective mass for all the cases considered,
stabilizing its high density behaviour.\\
An interesting aspect of the volume effects arises in
the relative baryonic populations, $n_B/n$, and leptonic populations,
$n_l/n$, as a function of the baryon density, figures 5a and 5b.
In figure 5a we study the change in the relative particle populations for
the ZM model, one can see that the finite volume
correlations enhance the production of hyperons as compared with the
point-like case. Furthermore the $\Lambda$ particles appear at a
relative density of $2.5$ (thin medium-dashed line), a bit earlier than
in the point-like description (thick medium-dashed line).
This could be expected since at high densities
the finite size correlations favours the creation of heavier electrically
neutral particles, decreasing the pressure of the system.
This last statement proves to be true provided the size of the hyperon is
similar to the size of the nucleons, as occurs in the present case. \\
In figure 5b we plot the relative particle population for the $QMCb$
treatment, a qualitative agreement with this figure is obtained when
the stability condition $QMCa$ is used. It can be seen that in the QMC model
the inclusion of finite volume correlations modifies very little the
relative presence of the hyperon $\Lambda$, which appears at lower
densities with respect to the predictions of the ZM model.\\
These results can be also compared with those obtained in 
ref.\cite{GLENDENN}, where baryons are considered as point-like.

\begin{figure}[ht]
\begin{tabular}{cc}
\begin{minipage}[t]{6.6cm}
\epsfig{file=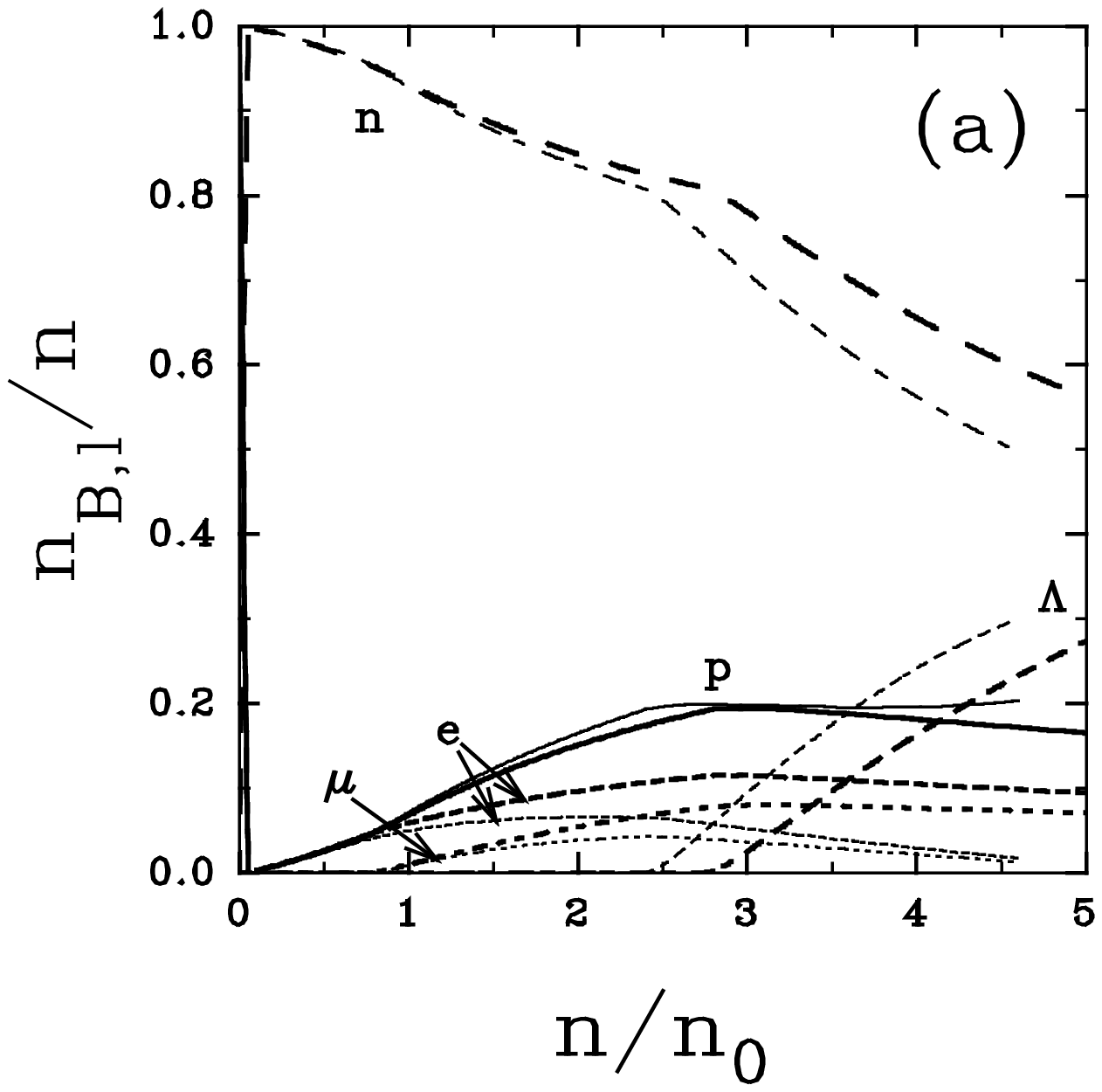,width=1.\textwidth}   
\end{minipage} &
\begin{minipage}[t]{6.6cm}
\epsfig{file=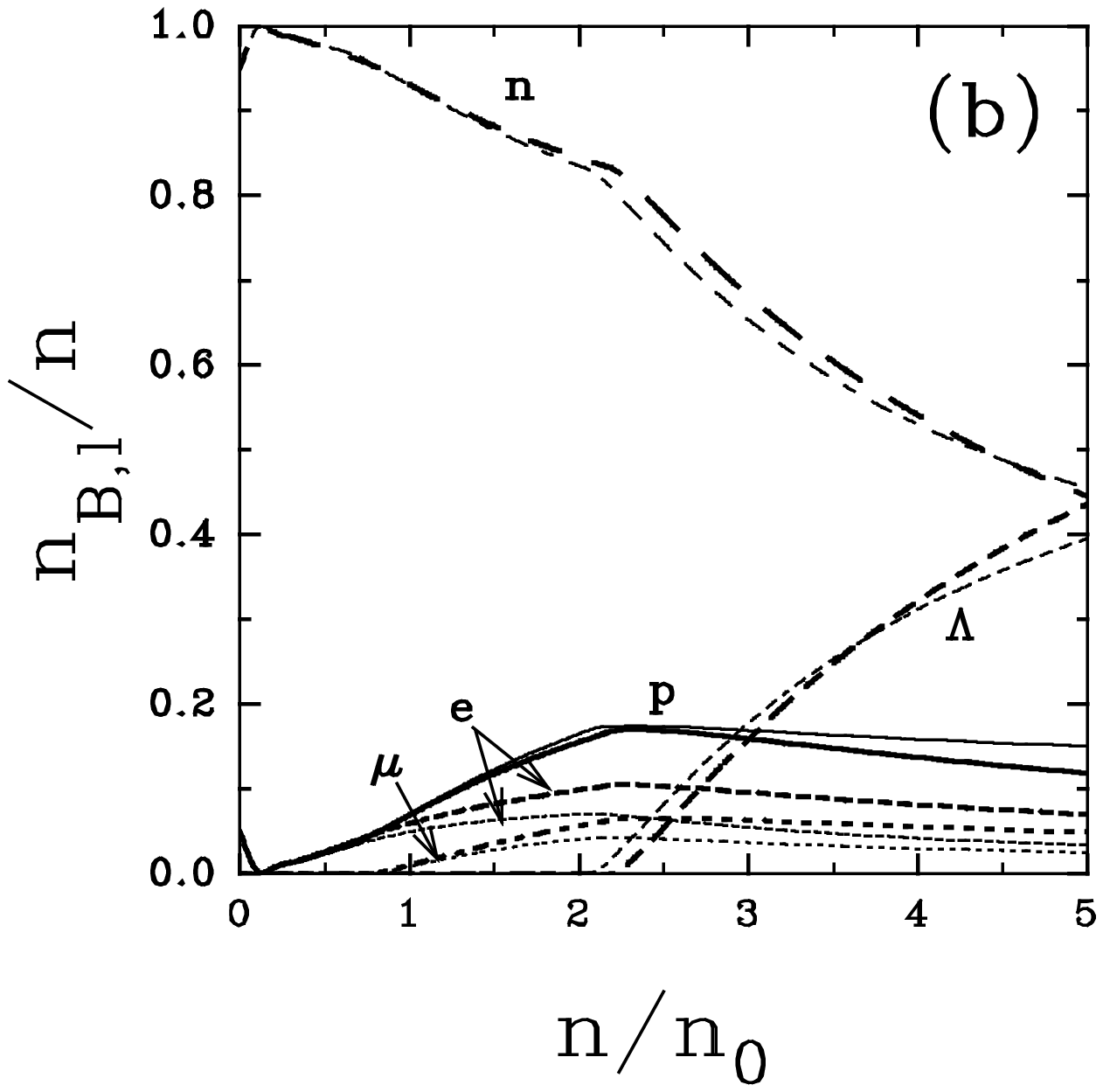,width=1.\textwidth}
\end{minipage}
\end{tabular}
\caption{\footnotesize{
Relative populations of baryons and leptons as functions of the
relative baryon density for the modified ZM model (a) and the QMC
model (b). In the last case only results for the equilibrium condition $QMCb$
are shown. Different line types are used for each particle.
For each class of particle thick (thin) lines represents $NC$ ($VC$)
calculations.}}
\end{figure}

\section{Conclusions}

We have introduced finite size volume correlations to extend the
study of matter properties at densities much higher than that of symmetric
nuclear matter at the saturation point. This has been performed through
a Van der Waals approach to describe hypermatter at zero temperature.
One of the advantages of the present approach consists in that we can
continue  to use the relativistic effective Lagrangians appropriate for
low baryonic density, because the excluded volume corrections appear
within the normalization of the baryon fields.\\
We have choosen two effective models to compare the effects of finite
volume correlations with respect to the more usual point-like descriptions,
these are the ZM and the QMC models. First we have enlarged the range of
applicability of Van der Waals volume correction to higher densities by
introducing a model (QMC) that avoids causality violation \cite{StoGre}-
\cite{SINGH},
since the hard core has been replaced with a compressible bag structure.
Second, although the QMC model deals with finite volume bags,
this property is usually lost in the calculation of thermodynamical properties.
This feature has been partially corrected in the results QMC-VC of this
work.\\ Near the normal saturation
density finite volume effects prove to be small for both the ZM and
the QMC models. At higher densities these repulsive short range effects
show up in a model dependent way. For the ZM model with rigid hadrons,
the pressure and the energy raise steeper than for the point-like prediction as  
the density gets closer to the touching limit. This favours
the production of $\Lambda$ hyperons in our example.\\
For the QMC model we have obtained that the contribution of finite effects
depends significantly on the bag equilibrium condition used, as explained in
the text. Using the standard condition the bags are not seriously compressed
with density and volume corrections can be appreciable in the EOS but not
in the change of relative particle occupation.\\
On the other side, if
the bags must equilibrate the external baryonic pressure, a seemingly
more physical condition, their respective radius suffer a faster reduction
when the density increases. Therefore  assuming the validity of this
approach, short range correlations could be neglected
in the EOS up to much higher densities, provided the compression effects
are included in the baryonic effective mass as the QMC model does. In fact
those physical aspects related to the size of the hadrons itself will be seriously
affected by the choise of the equilibrium conditions. \\
A general conclusion is that finite volume corrections are less
significative in the EOS when the medium density influence on the hadron
structure is properly taken into account inside the model prescriptions.

The production of hyperons is sensibly affected by excluded volume
correlations, decreasing its density threshold and enhancing its relative
population as density increases.

This studies on finite baryonic size effects on hyperon matter properties
aim to describe more properly the possible composition and structure
of neutron stars. However a more complete treatment must include the full
baryon octet.

\newpage
\renewcommand{\theequation}{\Alph{section}.\arabic{equation}}

%\appendix
\setcounter{section}{1}
\section*{Appendix}

In order to deduce the equilibrium condition proposed for case $QMCb$ we
consider only $\sigma$ and $\omega$ mesons. The inclusion of other mesons
is straigthforward.
The QMC lagrangian density with quark fields $\Psi_q(x)$
coupled to scalar $\sigma(x)$ and vector $\omega_{\mu}(x)$ neutral
mesons, is written as follows
\begin{eqnarray}
&{\cal L}_{QMC}(x) = \sum^3_{q=1}{\cal L}_q (x) + {\cal L}^0_{Mesons}(x) \ ,
& \\
&{\cal L}_q(x) = \left( {\cal L}^0_q(x) - B \right) \Theta_V
- \frac{1}{2} \bar{\Psi}_q(x) \Psi_q(x)
\Delta_S \ , &
\label{LAGR1}
\end{eqnarray}
Here $\Theta_V$ is the radial step function which schematically confine the
quarks inside the spherical bag.
The terms proportional to the surface delta function $\Delta_S$
ensures a zero net flux of quark current through the bag surface. In this
lagrangian  we have defined the following terms:

\begin{eqnarray}
{\cal L}^0_q(x)& =& \bar{\Psi}_q(x)
\left( i \gamma^\mu{\partial}_\mu - m_q + g_\sigma^q \sigma(x) -
g_\omega^q \gamma^\mu \omega_\mu(x) \right) \Psi_q(x),  \nonumber\\
{\cal L}^0_{Mesons}(x)& = & \frac{1}{2} [ \partial^\mu \sigma(x)
\partial_\mu \sigma(x) - m_\sigma^2 \sigma^2(x) ]
- \frac{1}{4} F^{\mu \nu}(x) F_{\mu \nu}(x)  \nonumber \\
&&+\frac{1}{2} m_\omega^2 \omega^\mu(x) \omega_\mu(x)  \,\,\, \nonumber
\label{LAGR2}
\end{eqnarray}
From this lagrangian the equations of motion of quarks and mesons can be
derived

\begin{eqnarray}
&\Theta_V (i \not \!\partial - m_q + g_{\sigma}^q \sigma - g_{\omega}^q
\not \! \omega) \Psi_q(x)=0, \label{EQUARK}&\\
& (\Box + m_{\sigma}^2) \sigma =
\sum^3_{q=1}g_{\sigma}^q \bar{\Psi}_q \Psi_q   \; \Theta_V, &\label{EQSCAL}
\\
&\partial_{\mu}F^{\mu \nu}+ m_{\omega}^2 \omega^{\nu}=
\sum^3_{q=1}g_{\omega}^q \bar{\Psi}_q \gamma^{\nu} \Psi_q  \; \Theta_V ,&
\label{EQVEC} \,\,\,
\end{eqnarray}
these equations must be completed with the boundary conditions at the bag
surface for the quark fields : $i {\mathbf \gamma \cdot n} \Psi_q(r=R) =
\Psi_q(r=R)$ and $i \bar{\Psi}_q(r=R) = \bar{\Psi}_q(r=R){\mathbf \gamma
\cdot n}$.

In the following equations summation over quark-flavor indices is assumed.

The energy-momentum tensor is evaluated by the canonical procedure giving
\begin{equation}
T_{QMC}^{\mu \nu}=\frac{1}{2} i \Theta_V (\bar{\Psi} \gamma^{\mu}
\stackrel {\leftrightarrow}{\partial^{\nu}} \Psi)+
\partial^{\mu}\sigma
\partial^{\nu}\sigma - F^{\mu \lambda} \partial^{\nu} \omega_{\lambda}-
g^{\mu \nu} {\cal L}
\end{equation}
inserting the explicit form of ${\cal L}$ and by taking its
divergence one obtains
\begin{eqnarray}
\partial_{\mu} T_{QMC}^{\mu \nu}
 &=& \frac{1}{2} i \Theta_V \partial_{\mu}(\bar{\Psi}_q \gamma^{\mu}
\stackrel{\leftrightarrow}{\partial^{\nu}} \Psi_q) +
\frac{1}{2} i n_{\mu} \Delta_S	(\bar{\Psi}_q \gamma^{\mu}
\stackrel{\leftrightarrow}{\partial^{\nu}} \Psi_q)
+ \frac{1}{2}\partial^{\nu}(\bar{\Psi_q}\Psi_q \Delta_S) \nonumber \\ && -
\Theta_V
\partial^{\nu}[\frac{1}{2}i\bar{\Psi}_q\stackrel{\leftrightarrow} {\not
\! \partial}\Psi_q + \bar{\Psi}_q (-m_q + g_{\sigma}^q \sigma - g_{\omega}^q
\not \! \omega)\Psi_q] \nonumber    \nonumber \\&& 
+ Bn^{\nu} \Delta_S - [\frac{1}{2}i\bar{\Psi}_q\stackrel{\leftrightarrow} 
{\not \!\partial}\Psi_q + \bar{\Psi}_q (-m_q + g_{\sigma}^q \sigma - 
g_{\omega}^q \not
\! \omega)\Psi_q] n^{\nu} \Delta_S \nonumber \\
&&-\partial_{\mu} F^{\mu \lambda} \partial^{\nu} \omega_{\lambda}
+\Box \sigma \partial^{\nu} \sigma + \frac{1}{2}
\partial^{\nu} [  m_{\sigma}^2 \sigma^2 +
m_{\omega}^2 \omega_{\lambda} \partial^{\nu} \omega^{\lambda}]\, , 
\end{eqnarray}

Using the equations of motion and rearranging terms one gets
\begin{eqnarray}
&\partial_{\mu} T_{QMC}^{\mu \nu} = -\Theta_V
\bar{\Psi}_q \partial^{\nu} (g_{\sigma}^q \sigma - g_{\omega}^q \not \!
\omega) \Psi
 + \frac{1}{2}i \Delta_S   n_{\mu} (\bar{\Psi}_q \gamma^{\mu}
\stackrel{\leftrightarrow}{\partial^{\nu}} \Psi) 
+ \frac{1}{2} \partial^\nu (\bar{\Psi}_q \Psi_q \Delta_S )
\nonumber &\\ & -
[\frac{1}{2} i \bar{\Psi}_q \stackrel{\leftrightarrow}{\not \!
\partial} \Psi_q  + \bar{\Psi}_q (-m_q + g_{\sigma}^q \sigma - g_{\omega}^q
\not\! \omega) \Psi_q ] n^{\nu} \Delta_S  
   \nonumber &\\ &+ B n^{\nu} \Delta_S +(\Box
+ m_{\sigma}^2)\sigma \;
\partial^{\nu}\sigma - (\partial{\mu} F^{\mu \lambda} +
m_{\omega}^2 \omega^{\lambda})
\partial^{\nu} \omega_{\lambda} &
\end{eqnarray}

\noindent
the two terms in the squared bracket are canceled by using (\ref{EQUARK}).
If we denote by $\tilde{\Theta}_V$ the radial step function complementary to
$\Theta_V$, introducing the separation
$1=\Theta_V +\tilde{\Theta}_V$ in the last line and reagrouping terms
\begin{eqnarray}
\partial_{\mu} T_{QMC}^{\mu \nu}& =&
 \Theta_V [(\Box+ m_{\sigma}^2)\sigma -
 g_{\sigma}^q \bar{\Psi}_q \Psi_q]\partial^{\nu}\sigma \nonumber \\ &&
 - \Theta_V [(\partial{\mu} F^{\mu \lambda} +
m_{\omega}^2 \omega^{\lambda})- \bar{\Psi}_q g_{\omega}^q \gamma^\lambda
\Psi_q] \partial^{\nu} \omega_{\lambda} + B n^\nu \Delta_S  \nonumber
\\ && + \frac{1}{2} \Delta_S  [ i n_{\mu} (\bar{\Psi}_q \gamma^{\mu}
\stackrel{\leftrightarrow}{\partial^{\nu}} \Psi_q)] \nonumber  +
\frac{1}{2} \partial^\nu (\bar{\Psi}_q \Psi_q \Delta_S ) \nonumber \\
&&+[(\Box + m_{\sigma}^2)\sigma \;
\partial^{\nu}\sigma - (\partial{\mu} F^{\mu \lambda} +
m_{\omega}^2 \omega^{\lambda})
 \partial^{\nu} \omega_{\lambda}]\tilde{\Theta}_V \,\,\, .
\end{eqnarray}
The first two terms on the right hand side are identically zero by
virtue of Eqs. (\ref{EQSCAL}) and (\ref{EQVEC}) respectively, therefore it
becomes

\begin{eqnarray}
\partial_{\mu} T_{QMC}^{\mu \nu}& =&
\Delta_S  ( B n^\nu + \frac{1}{2} i n_{\mu} \bar{\Psi}_q \gamma^{\mu}
\stackrel{\leftrightarrow}{\partial^{\nu}} \Psi_q) \nonumber \\ &&
+\frac{1}{2} \partial^\nu (\bar{\Psi}_q \Psi_q \Delta_S ) + \partial_{\mu}
T_{Mesons}^{\mu \nu} \tilde{\Theta}_V
\end{eqnarray}

where
\begin{eqnarray}
\partial_{\mu} T_{Mesons}^{\mu \nu}&=&
(\Box + m_{\sigma}^2)\sigma \;
\partial^{\nu}\sigma
- (\partial{\mu} F^{\mu \lambda} + m_{\omega}^2
\omega^{\lambda})
 \partial^{\nu} \omega_{\lambda}
\end{eqnarray}

In this way we can rewrite the above expression as follows
\begin{eqnarray}
n_\nu \partial_{\mu} T^{\mu \nu}_{QMC} & = & n_\nu \partial_{\mu}
T^{\mu \nu}_{bag} + n_\nu \partial_{\mu} T^{\mu \nu}_{Mesons}
{\tilde{\Theta}_V} \label{EQUIL}\\
{\mathrm with}&& \; \; \; \; \; \; \; \; \; \;\nonumber\\
 n_\nu \partial_{\mu} T^{\mu \nu}_{bag} & = &
(-P_D+B) \Delta_S  + \frac{1}{2} n \cdot \partial (\bar{\Psi}_q \Psi_q
\Delta_S )
\end{eqnarray}

\noindent
where $P_D=-1/2 n \cdot  \partial({\bar{\Psi}_q}\Psi_q)|_{surface}$ is the
pressure exerted by the quarks on the bag surface.
The standard QMC equilibrium condition can be derived by requiring to be
zero the first term of Eq.(\ref{EQUIL}). Particularly, for the QMC model the
second term of this equation is strictly zero, which is due to the fact that
in the QMC lagrangian there is a static bag surrounded by free mesons. As
the baryon density increases this picture is not adequate, since
at high densities the mesons are far from the free field approximation.
Although mesons are in the whole space, i.e. outside as well as
inside the bags, the $\tilde{\Theta}_V$ in Eq.(\ref{EQUIL}) is
coming from a cancellation of identical terms as it has been shown.
The fact that mesons are also inside the bag is reflected on the
baryon density dependence of $m_q$, $x_q$, $R$, the quark wave
functions, etc.

On the other hand, from the effective Lagrangian \ref{LAGRAL} we have
$T^{\mu \nu}_{hadron} = T^{\mu \nu}_{fm}+T^{\mu \nu}_0 $, where $
T^{\mu \nu}_{fm} $ comes from the free meson sector and $ T^{\mu
\nu}_0 $ is the contribution from baryons coupled to mesons.
Energy-momentum conservation implies $\partial_{\mu} T^{\mu
\nu}_{hadron}=0$. Explicit evaluation shows that $\partial_{\mu}
T^{\mu \nu}_{fm}=\partial_{\mu} T^{\mu \nu}_{Mesons}$.
Consequently, in this limit the second term of Eq.(\ref{EQUIL}) can be
replaced by $- \partial_{\mu} T^{\mu \nu}_0$. Multiplying the
resulting equation by $n_{\nu}$ and integrating on the external
bag volume we obtain
\begin{equation}
4 \pi R^2 (-P_D + B) = \int dS n_{\mu} n_{\nu} T^{\mu \nu}_0 \, .
\label{CONSERV}
\end{equation}
For homogeneous hadronic matter $T^{\mu \nu}_0$ can be assumed as
constant on the bag surface, obtaining the following equation
\begin{equation}
-P_D + B =  n_{\mu} n_{\nu} T^{\mu \nu}_0 (r=R)
\end{equation}
where $n_{\lambda}$ is a (space-like) radial unit vector.
Therefore, we get
\begin{equation}
-P_D + B = {{1}\over{3}} \sum_{i,j} T^{i j}_0
\end{equation}
In the case of homogeneous static matter the space sector of the
relativistic energy-momentum tensor is diagonal \cite{LANDAU},
which implies that
\begin{equation}
-P_D + B = {{1}\over{3}} \sum_{i}^3 T^{i i}_0 = -P_0
\end{equation}
where $P_0$ is a pressure related to $T_0$. In this form we have
obtained the equilibrium condition of case $QMCb$.

\end{document}